\documentclass[10pt,letterpaper]{article}
\usepackage{amsmath}

 \evensidemargin0in \oddsidemargin0in \topmargin10pt
\begin{document}

\title{Relation between Fresnel transform of input light field and Radon
transform of Wigner function of the field}
\author{Hong-yi Fan and Li-yun Hu$^{\ast }$ \\
{\small Department of Physics, Shanghai Jiao Tong University, }\\
{\small \ Shanghai, 200030, China}\\
{\small *Corresponding author: hlyun@sjtu.edu.cn; hlyun@sjtu.org}}
\maketitle

\begin{abstract}
{\small We prove a new theorem about the relationship between optical field
Wigner function's Radon transform and optical Fresnel transform of the
field, i.e., when an input field }$\psi \left( x^{\prime }\right) ${\small \
propagates through an optical }$\left[ D\left( -B\right) \left( -C\right) A%
\right] ${\small \ system, the energy density of the output field is equal
to the Radon transform of the Wigner function of the input field, where the
Radon transform parameters are }$D,B.${\small \ We prove this theorem in
both spatial-domain and frequency-domain.}
\end{abstract}

In optical communication theory every signal or image can be uniquely and
indirectly described by a Wigner distribution function (WDF) \cite{1,2,3}.
The WDF (or named Wigner transform) of an optical signal field $\psi \left(
x^{\prime }\right) $ is defined as
\begin{equation}
W_{\psi }(\nu ^{\prime },x^{\prime })=\int_{-\infty }^{+\infty }\frac{du}{%
2\pi }e^{i\nu ^{\prime }u}\psi ^{\ast }\left( x^{\prime }+\frac{u}{2}\right)
\psi \left( x^{\prime }-\frac{u}{2}\right) ,  \label{1}
\end{equation}%
$W_{\psi }(\nu ^{\prime },x^{\prime })$ involves both spatial distribution
information and space-frequency distribution information of the signal. $\nu
$ is named space frequency. $W_{\psi }(\nu ^{\prime },x^{\prime })$ is said
to be bilinear in the signal because the signal enters twice in its
definition. The WDF undergoes certain variations if something happens to the
signal. For examples, passage through a lens corresponds to a vertical
shearing of the WDF, propagation in free space means a horizontal shearing
of the WDF \cite{4}. However, the WDF preserves space and space frequency
marginal properties of any signal,
\begin{eqnarray}
\int_{-\infty }^{+\infty }d\nu ^{\prime }W_{\psi }(\nu ^{\prime },x^{\prime
}) &=&|\psi \left( x^{\prime }\right) |^{2},  \label{2} \\
\int_{-\infty }^{+\infty }dx^{\prime }W_{\psi }(\nu ^{\prime },x^{\prime })
&=&|\tilde{\psi}\left( \nu ^{\prime }\right) |^{2},\text{ }  \label{3}
\end{eqnarray}%
where $\tilde{\psi}\left( \nu \right) =\int_{-\infty }^{+\infty }\frac{dx}{%
\sqrt{2\pi }}\psi \left( x\right) e^{ix\nu }.$ If one wants to reconstruct
the Wigner function by using various probability distribution, obviously the
position density $|\psi \left( x^{\prime }\right) |^{2}$ and the
space-frequency density $|\tilde{\psi}\left( \nu ^{\prime }\right) |^{2}$
were not enough, so the Radon transform \cite{5,6} of the Wigner function is
introduced \cite{7},%
\begin{equation}
R\left( x\right) \equiv \iint\limits_{-\infty }^{\infty }dx^{\prime }d\nu
^{\prime }\delta \left( x-Dx^{\prime }+B\nu ^{\prime }\right) W_{\psi }(\nu
^{\prime },x^{\prime }),  \label{4}
\end{equation}%
$R\left( x\right) $ is also a probability distribution along an infinitely
thin phase space strip denoted by the real parameters $D,B$. The inverse
relation of (\ref{4}) is the foundation of optical tomographic imaging
techniques (the techniques derive two-dimensional data from a
three-dimensional object to obtain a slice image of the internal structure
and thus have the ability to peer inside the object noninvasively.)

On the other hand, an optical system can be analyzed by using either
diffraction theory or ray optics. It is worth setting up a connection
between ray transfer matrix and diffraction theory, i.e., the diffraction
integration written in terms of ray transfer matrix---named Collins'
diffraction integral formula \cite{8,9,10} which describes the propagation
of a general beam $\psi \left( x^{\prime }\right) $ through an $\left(
ABCD\right) $ optical paraxial system,
\begin{equation}
\phi \left( x\right) =\int_{-\infty }^{\infty }\mathcal{K}\left( x,x^{\prime
}\right) \psi \left( x^{\prime }\right) dx^{\prime },  \label{5}
\end{equation}%
where $\phi \left( x\right) $ is the output field, $AD-BC=1,$ and $\mathcal{K%
}\left( x,x^{\prime }\right) $ is the integral kernel%
\begin{equation}
\mathcal{K}\left( x,x^{\prime }\right) =\frac{1}{\sqrt{2\pi iB}}\exp \left[
\frac{i}{2B}\left( Ax^{\prime 2}-2x^{\prime }x+Dx^{2}\right) \right] .
\label{6}
\end{equation}%
If the energy in the initial beam is normalized, $\int_{-\infty }^{+\infty
}\left\vert \psi \left( x^{\prime }\right) \right\vert ^{2}dx^{\prime }=1,$
then the output beam is normalized too, $\int_{-\infty }^{+\infty
}\left\vert \phi \left( x\right) \right\vert ^{2}dx=1.$ Clearly, if the $%
\left[ ABCD\right] $ system is changed to $\left[ D\left( -B\right) \left(
-C\right) A\right] $ system, then Eq. (\ref{5}) should read%
\begin{equation}
\phi \left( x\right) =\int_{-\infty }^{\infty }\mathcal{\vec{K}}\left(
x,x^{\prime }\right) \psi \left( x^{\prime }\right) dx^{\prime },  \label{7}
\end{equation}%
where the integral kernel $\mathcal{\vec{K}}\left( x,x^{\prime }\right) $ is%
\begin{equation}
\mathcal{\vec{K}}\left( x,x^{\prime }\right) =\frac{1}{\sqrt{-2\pi iB}}\exp %
\left[ \frac{-i}{2B}\left( Dx^{\prime 2}-2x^{\prime }x+Ax^{2}\right) \right]
.  \label{8}
\end{equation}

In this Letter, we shall reveal the following\textbf{\ theorem}:

\textbf{When an input field }$\psi \left( x^{\prime }\right) $\textbf{\
propagates through an optical }$\left[ D\left( -B\right) \left( -C\right) A%
\right] $\textbf{\ system, the energy density of the output field }$\phi
\left( x\right) $\textbf{\ is equal to the Radon transform of the Wigner
function of the input field, where the Radon transform parameters are }$D,B$%
\textbf{.} The proof is demonstrated as follows.

With the use of Dirac $\delta $-function, we can re-express the Wigner
function of input field $\psi \left( x^{\prime }\right) $ in Eq. (\ref{1}) as%
\begin{eqnarray}
W_{\psi }(\nu ^{\prime },x^{\prime }) &=&\iint\limits_{-\infty }^{\infty
}dx^{\prime \prime }dx^{\prime \prime \prime }\int_{-\infty }^{+\infty }%
\frac{du}{2\pi }e^{i\nu ^{\prime }u}\psi ^{\ast }\left( x^{\prime \prime
}\right) \psi \left( x^{\prime \prime \prime }\right) \delta \left(
x^{\prime \prime }-x^{\prime }-\frac{u}{2}\right) \delta \left( x^{\prime }-%
\frac{u}{2}-x^{\prime \prime \prime }\right)   \notag \\
&=&\frac{1}{\pi }\iint\limits_{-\infty }^{\infty }dx^{\prime \prime
}dx^{\prime \prime \prime }\psi ^{\ast }\left( x^{\prime \prime }\right)
\psi \left( x^{\prime \prime \prime }\right) e^{i2\nu ^{\prime }\left(
x^{\prime \prime }-x^{\prime }\right) }\delta \left( 2x^{\prime }-x^{\prime
\prime }-x^{\prime \prime \prime }\right) .  \label{9}
\end{eqnarray}%
Substituting (\ref{9}) into (\ref{4}) we rewrite the Radon transform of $%
W_{\psi }(\nu ^{\prime },x^{\prime })$ as%
\begin{eqnarray}
R\left( x\right)  &=&\frac{1}{\pi }\iint\limits_{-\infty }^{\infty
}dx^{\prime }d\nu ^{\prime }\delta \left( x-Dx^{\prime }+B\nu ^{\prime
}\right) \iint\limits_{-\infty }^{\infty }dx^{\prime \prime }dx^{\prime
\prime \prime }\psi ^{\ast }\left( x^{\prime \prime }\right) \psi \left(
x^{\prime \prime \prime }\right) e^{i2\nu ^{\prime }\left( x^{\prime \prime
}-x^{\prime }\right) }\delta \left( 2x^{\prime }-x^{\prime \prime
}-x^{\prime \prime \prime }\right)   \notag \\
&=&\frac{1}{2\pi B}\iint\limits_{-\infty }^{\infty }dx^{\prime \prime
}dx^{\prime \prime \prime }\psi ^{\ast }\left( x^{\prime \prime }\right)
\psi \left( x^{\prime \prime \prime }\right) \exp \left\{ \frac{i}{2B}%
[D\left( x^{\prime \prime 2}-x^{\prime \prime \prime 2}\right) -2x\left(
x^{\prime \prime }-x^{\prime \prime \prime }\right) \right\} .  \label{10}
\end{eqnarray}%
On the other hand, when the beam $\psi \left( x^{\prime }\right) $
propagates through the $\left[ D\left( -B\right) \left( -C\right) A\right] $
optical system, according to the Fresnel integration (\ref{7})-(\ref{8}), we
have%
\begin{eqnarray}
\left\vert \phi \left( x\right) \right\vert ^{2} &=&\int_{-\infty }^{\infty
}dx^{\prime }\mathcal{\vec{K}}\left( x,x^{\prime }\right) \psi \left(
x^{\prime }\right) \int_{-\infty }^{\infty }dx^{\prime \prime }\psi ^{\ast
}\left( x^{\prime \prime }\right) \mathcal{\vec{K}}^{\ast }\left(
x,x^{\prime \prime }\right)   \notag \\
&=&\frac{1}{2\pi B}\int_{-\infty }^{\infty }dx^{\prime }\exp \left[ \frac{-i%
}{2B}\left( Dx^{\prime 2}-2x^{\prime }x+Ax^{2}\right) \right] \psi \left(
x^{\prime }\right)   \notag \\
&&\times \int_{-\infty }^{\infty }dx^{\prime \prime }\exp \left[ \frac{i}{2B}%
\left( Dx^{\prime \prime 2}-2x^{\prime \prime }x+Ax^{2}\right) \right] \psi
^{\ast }\left( x^{\prime \prime }\right)   \notag \\
&=&\frac{1}{2\pi B}\iint\limits_{-\infty }^{\infty }dx^{\prime }dx^{\prime
\prime }\psi \left( x^{\prime }\right) \psi ^{\ast }\left( x^{\prime \prime
}\right) \exp \left\{ \frac{i}{2B}\left[ D\left( x^{\prime \prime
2}-x^{\prime 2}\right) +2x\left( x^{\prime }-x^{\prime \prime }\right) %
\right] \right\} ,  \label{11}
\end{eqnarray}%
which is the same as $R\left( x\right) $ in (\ref{10})$.$ So combining (\ref%
{4}), (\ref{10})-(\ref{11}), (\ref{7}) and (\ref{8}) we reach the conclusion%
\begin{equation}
\left\vert \frac{1}{\sqrt{-2\pi iB}}\int_{-\infty }^{\infty }\exp \left[
\frac{-i}{2B}\left( Dx^{\prime 2}-2x^{\prime }x+Ax^{2}\right) \right] \psi
\left( x^{\prime }\right) dx^{\prime }\right\vert ^{2}=\iint\limits_{-\infty
}^{\infty }dx^{\prime }d\nu ^{\prime }\delta \left( x-Dx^{\prime }+B\nu
^{\prime }\right) W_{\psi }(\nu ^{\prime },x^{\prime }),  \label{12}
\end{equation}%
where $DA-BC=1$. The physical meaning of Eq. (\ref{12}) is: when an input
field propagates through an optical $\left[ D\left( -B\right) \left(
-C\right) A\right] $ system, the energy density of the output field is equal
to the Radon transform of the Wigner function of the input field. So far as
our knowledge is concerned, this conclusion seems new.

Eq. (\ref{8}) is the relationship between the input amplitude and output one
in spatial-domain. Now we turn the above discussion to the case of
space-frequency domain.

For a $\left[ D\left( -B\right) \left( -C\right) A\right] $ optical system,
the Collins' diffraction integral formula in space-frequency (angle spectrum
\cite{11}) domain is \cite{12}%
\begin{equation}
\widetilde{\phi }\left( v\right) =\int_{-\infty }^{\infty }\widetilde{%
\mathcal{K}}\left( {\nu },\nu ^{\prime }\right) \widetilde{\psi }\left( {\nu
}^{\prime }\right) d{\nu }^{\prime },  \label{13}
\end{equation}%
where the kernel is
\begin{equation}
\widetilde{\mathcal{K}}\left( {\nu },{\nu }^{\prime }\right) =\frac{1}{\sqrt{%
2\pi iC}}\exp \left[ \frac{i}{2C}\left( D{\nu }^{2}-2{\nu }^{\prime }{\nu }+A%
{\nu }^{\prime 2}\right) \right] .  \label{14}
\end{equation}%
On the other hand, the Wigner function expressed in terms of the
space-frequency field is

\begin{equation}
W_{\tilde{\psi}}(\nu ^{\prime },x^{\prime })=\int_{-\infty }^{+\infty }\frac{%
ds}{2\pi }e^{-ix^{\prime }s}\tilde{\psi}^{\ast }\left( \nu ^{\prime }+\frac{s%
}{2}\right) \tilde{\psi}\left( \nu ^{\prime }-\frac{s}{2}\right) .
\label{15}
\end{equation}%
In space-frequency domain the Radon transform is along an infinitely thin
phase space strip denoted by the real parameters $A,C$,%
\begin{equation}
R\left( \nu \right) \equiv \iint\limits_{-\infty }^{\infty }dx^{\prime
}dp^{\prime }\delta \left( {\nu }-A{\nu }^{\prime }+Cx^{\prime }\right) W_{%
\tilde{\psi}}({\nu }^{\prime },x^{\prime })  \label{16}
\end{equation}%
rather than the parameters $D,B$ of the Radon transform (\ref{4}) in
spatial-domain. We want to examine if the above conclusion still holds in
the space-frequency domain. For this propose, we rewrite (\ref{15}) as
\begin{eqnarray}
W_{\tilde{\psi}}(\nu ^{\prime },x^{\prime }) &=&\iint\limits_{-\infty
}^{\infty }d{\nu }^{\prime \prime }d{\nu }^{\prime \prime \prime
}\int_{-\infty }^{+\infty }\frac{ds}{2\pi }e^{-ix^{\prime }s}\tilde{\psi}%
^{\ast }\left( {\nu }^{\prime \prime }\right) \tilde{\psi}\left( {\nu }%
^{\prime \prime \prime }\right) \delta \left( \nu ^{\prime \prime }-\nu
^{\prime }-\frac{s}{2}\right) \delta \left( \nu ^{\prime }-\frac{s}{2}-\nu
^{\prime \prime \prime }\right)   \notag \\
&=&\iint\limits_{-\infty }^{\infty }d{\nu }^{\prime \prime }d{\nu }^{\prime
\prime \prime }\tilde{\psi}^{\ast }\left( {\nu }^{\prime \prime }\right)
\tilde{\psi}\left( {\nu }^{\prime \prime \prime }\right) e^{-i2x^{\prime
}\left( {\nu }^{\prime \prime }-{\nu }^{\prime }\right) }\delta \left( 2{\nu
}^{\prime }-{\nu }^{\prime \prime }-{\nu }^{\prime \prime \prime }\right)
\label{17}
\end{eqnarray}%
Then we substitute (\ref{17}) into (\ref{16}), the result is%
\begin{eqnarray}
R\left( \nu \right)  &=&\frac{1}{\pi }\iint\limits_{-\infty }^{\infty
}dx^{\prime }d{\nu }^{\prime }\delta \left( {\nu }-A{\nu }^{\prime
}+Cx^{\prime }\right) \iint\limits_{-\infty }^{\infty }d{\nu }^{\prime
\prime }d{\nu }^{\prime \prime \prime }\tilde{\psi}^{\ast }\left( {\nu }%
^{\prime \prime }\right) \tilde{\psi}\left( {\nu }^{\prime \prime \prime
}\right) e^{-i2x^{\prime }\left( {\nu }^{\prime \prime }-{\nu }^{\prime
}\right) }\delta \left( 2{\nu }^{\prime }-{\nu }^{\prime \prime }-{\nu }%
^{\prime \prime \prime }\right)   \notag \\
&=&\frac{1}{2\pi C}\iint\limits_{-\infty }^{\infty }d{\nu }^{\prime \prime }d%
{\nu }^{\prime \prime \prime }\tilde{\psi}^{\ast }\left( {\nu }^{\prime
\prime }\right) \tilde{\psi}\left( {\nu }^{\prime \prime \prime }\right)
\exp \left\{ -\frac{i}{2C}\left[ A\left( {\nu }^{\prime \prime 2}-{\nu }%
^{\prime \prime \prime 2}\right) -2{\nu }\left( {\nu }^{\prime \prime }-{\nu
}^{\prime \prime \prime }\right) \right] \right\} .  \label{18}
\end{eqnarray}%
On the other hand, from (\ref{13}) we calculate
\begin{eqnarray}
|\tilde{\phi}\left( {\nu }\right) |^{2} &=&\frac{1}{2\pi C}\int_{-\infty
}^{+\infty }\tilde{\psi}\left( {\nu }^{\prime }\right) \exp \left[ \frac{i}{%
2C}\left( D{\nu }^{2}-2{\nu }^{\prime }{\nu }+A{\nu }^{\prime 2}\right) %
\right] d{\nu }^{\prime }  \notag \\
&&\times \int_{-\infty }^{+\infty }\tilde{\psi}^{\ast }\left( {\nu }^{\prime
\prime }\right) \exp \left[ \frac{-i}{2C}\left( D{\nu }^{2}-2{\nu }^{\prime
\prime }{\nu }+A{\nu }^{\prime \prime 2}\right) \right] d{\nu }^{\prime
\prime }  \notag \\
&=&\frac{1}{2\pi C}\int \int_{-\infty }^{+\infty }d{\nu }^{\prime }d{\nu }%
^{\prime \prime }\tilde{\psi}\left( {\nu }^{\prime }\right) \tilde{\psi}%
^{\ast }\left( {\nu }^{\prime \prime }\right) \exp \left[ -\frac{i}{2C}%
\left( A\left( {\nu }^{\prime \prime 2}-{\nu }^{\prime 2}\right) -2{\nu }%
\left( {\nu }^{\prime \prime }-{\nu }^{\prime }\right) \right) \right] .
\label{19}
\end{eqnarray}%
Comparing Eq. (\ref{19}) with Eq. (\ref{18}) and using (\ref{13})-(\ref{14})
we see%
\begin{equation}
\left\vert \frac{1}{\sqrt{2\pi iC}}\int \exp \left[ \frac{i}{2C}\left( D{\nu
}^{2}-2{\nu }^{\prime }{\nu }+A{\nu }^{\prime 2}\right) \right] \tilde{\psi}%
\left( {\nu }^{\prime }\right) d{\nu }^{\prime }\right\vert
^{2}=\iint\limits_{-\infty }^{\infty }dx^{\prime }dp^{\prime }\delta \left( {%
\nu }-A{\nu }^{\prime }+Cx^{\prime }\right) W_{\tilde{\psi}}({\nu }^{\prime
},x^{\prime }),  \label{20}
\end{equation}%
this is the theorem expressed in space-frequency domain.

We now take an example to confirm the theorem. When the input field is
described by a Gaussian function (Gaussian chirplet)
\begin{equation}
\psi _{0}\left( x^{\prime }\right) \equiv \sqrt[4]{\frac{\epsilon }{\pi }}%
\exp \left\{ -\left( \epsilon -i\beta \right) \frac{x^{\prime 2}}{2}\right\}
,\text{ }\epsilon >0,\text{\ }  \label{21}
\end{equation}%
which depicts a Gaussian windowed linear chirp signal \cite{13}. Using Eq.(%
\ref{1}) we obtain its Wigner function%
\begin{eqnarray}
W_{\psi _{0}}(\nu ^{\prime },x^{\prime }) &=&e^{-\epsilon x^{\prime }{}^{2}}%
\sqrt{\frac{\epsilon }{\pi }}\int_{-\infty }^{+\infty }\frac{du}{2\pi }e^{-%
\frac{1}{4}\epsilon u^{2}+i\left( \nu ^{\prime }-\beta x^{\prime }\right) u}
\notag \\
&=&\frac{1}{\pi }\exp \left\{ -\left[ \epsilon x^{\prime }{}^{2}+\frac{1}{%
\epsilon }\left( \nu ^{\prime }-\beta x^{\prime }\right) ^{2}\right]
\right\} ,  \label{22}
\end{eqnarray}%
which shows that the energy of Gaussian chirplet is concentrated at $\nu
^{\prime }=\beta x^{\prime }.$ According to Eq. (\ref{4}), the Radon
transform of (\ref{22}) with the parameters $D,B$ is%
\begin{eqnarray}
&&\iint\limits_{-\infty }^{\infty }dx^{\prime }d\nu ^{\prime }\delta \left(
x-Dx^{\prime }+B\nu ^{\prime }\right) W_{\psi _{0}}(\nu ^{\prime },x^{\prime
})  \notag \\
&=&\frac{1}{\pi D}\int_{-\infty }^{+\infty }d\nu ^{\prime }\exp \left\{ -%
\frac{\epsilon }{D^{2}}\left( x+B\nu ^{\prime }\right) {}^{2}-\frac{1}{%
\epsilon D^{2}}\left[ \left( D-\beta B\right) \nu ^{\prime }-\beta x\right]
^{2}\right\}  \notag \\
&=&\sqrt{\frac{\epsilon /\pi }{\left( D-B\beta \right) ^{2}+B^{2}\epsilon
^{2}}}\exp \left[ \frac{-\epsilon x^{2}}{\left( D-B\beta \right)
^{2}+B^{2}\epsilon ^{2}}\right] \equiv R_{0}\left( x\right) .  \label{23}
\end{eqnarray}%
On the other hand, according to (\ref{7}) and (\ref{8}), the Fresnel
transform of input field $W_{\psi _{0}}(\nu ^{\prime },x^{\prime })$ through
a $\left[ D\left( -B\right) \left( -C\right) A\right] $ optical system is
given by%
\begin{eqnarray}
&&\frac{1}{\sqrt{-2\pi iB}}\sqrt[4]{\frac{\epsilon }{\pi }}\int_{-\infty
}^{\infty }dx^{\prime }\exp \left[ \frac{-i}{2B}\left( Dx^{\prime
2}-2x^{\prime }x+Ax^{2}\right) -\left( \epsilon -i\beta \right) \frac{%
x^{\prime 2}}{2}\right]  \notag \\
&=&\sqrt[4]{\frac{\epsilon }{\pi }}\frac{1}{\sqrt{-i}}e^{-\frac{iA}{2B}x^{2}}%
\sqrt{\frac{1}{\epsilon B-i\left( \beta B-D\right) }}\exp \left[ \frac{-x^{2}%
}{2B\left( B\epsilon +iD-i\beta B\right) }\right] \left. \equiv \right. \phi
_{0}\left( x\right) .  \label{24}
\end{eqnarray}%
Comparing (\ref{24}) with (\ref{23}) and noticing%
\begin{equation}
\frac{1}{2B\left( B\epsilon +iD-i\beta B\right) }+\frac{1}{2B\left(
B\epsilon -iD+i\beta B\right) }=\allowbreak \frac{\epsilon }{\left( D-B\beta
\right) ^{2}+B^{2}\epsilon ^{2}}  \label{25}
\end{equation}%
we see%
\begin{equation}
\left\vert \phi _{0}\left( x\right) \right\vert ^{2}=R_{0}\left( x\right) ,
\label{26}
\end{equation}%
as expected.

In summary, we have derived a new theorem governing the connection between
optical field's Fresnel transform and its Wigner function's Radon
transformation, since both the Wigner function and the Fresnel transform are
widely used in optical propagation, we hope this theorem would have new
applications in the analysis of optical communication and optical
tomography. For the application of Wigner function in deriving the
quantum-mechanical photocount formula, we refer to \cite{14}.

This work was supported by the National Natural Science Foundation of China
under grant 10775097.


\begin{thebibliography}{99}
\bibitem{1} E. P. Wigner, "On the quantum correction for thermodynamic
equilibrium", Phys. Rev. \textbf{40}, 749 (1932).

\bibitem{2} M. j. Bastiaans, "The Wigner distribution function applied to
optical signals and systems", Opt. Commun. \textbf{25,} 26 (1978).

\bibitem{3} M. j. Bastiaans, "Wigner distribution function and its
application to first-order optics", J. Opt. Soc. Am. \textbf{69}, 1710
(1979).

\bibitem{4} A. W. Lohmann, \textquotedblleft Image rotation, Wigner
rotation, and the fractional Fourier transform,\textquotedblright\ J. Opt.
Soc. Am. A \textbf{10}, 2181--2186 (1993).

\bibitem{5} J. Radon, "Uber die Bestimmung von Funktionen Durch Ihre
Integralwerte Langs Gewisser Mannigfaltigkeiten", Ber. Verh. Saechs. Akad.
Wiss. Leipzig Math.Phys.K1. \textbf{69}, 262-267, (1917).

\bibitem{6} Y. Zhang, B. Gu, B. Dong, and G. Yang, "Optical implementations
of the Radon--Wigner display for one-dimensional signals," Opt. Lett. 23,
1126-1128 (1998).

\bibitem{7} For quantum states' reconstruction, see e.g., Wolfgang P.
Schleich, \textit{Quantum Optics in Phase Space}, (Wiley-VCH, Birlin, 2001)
and references therein

\bibitem{8} S. A. Collins, "Lens-system diffraction integral written in
terms of matrix optic", J. Opt. Soc. Am. A \textbf{60}, 1168-1177 (1970).

\bibitem{9} J. A. Arnaud, "Mode Coupling in First-Order Optics," J. Opt.
Soc. Am. \textbf{61}, 751-758 (1971).

\bibitem{10} H. Fan and H. Lu, "Collins diffraction formula studied in
quantum optics," Opt. Lett. \textbf{31}, 2622-2624 (2006).

\bibitem{11} J. W. Goodman, \textit{Introduction to Fourier Optics}, (Mc
Graw-Hill, 1968).

\bibitem{12} Z. Liu, X. Xiu and D. Fan, "Collins formula in frequency-domain
and fractional Fourier transforms", Opt. Commun. \textbf{155}, 7 (1998).

\bibitem{13} A. Papoulis, \textit{Signal Analysis,} (McGraw-Hill, New York,
1977).

\bibitem{14} Hong-yi Fan and Li-yun Hu, "Two quantum-mechanical photocount
formulas", Opt. Lett. \textbf{33}, 443-445 (2008).
\end{thebibliography}
\end{document}